# Does the Dirac Cone Exist in Silicene on Metal Substrates?

Ruge Quhe,[1,3,4,5] Yakun Yuan,[1] Jiaxin Zheng,[1,3] Yangyang Wang,[1] Zeyuan Ni,[1] Junjie Shi,[1] Dapeng Yu,[1,2] Jinbo Yang,[1,2]* Jing Lu,[1,2]*

[1]State Key Laboratory of Mesoscopic Physics and Department of Physics, Peking University, Beijing 100871, P. R. China

[2]Collaborative Innovation Center of Quantum Matter, Beijing 100871, P. R. China

[3]Academy for Advanced Interdisciplinary Studies, Peking University, Beijing 100871, P. R. China

[4]Department of Chemistry and Applied Biosciences, Eidgenössische Technische Hochschule Zürich, CH-8093 Zurich, Switzerland

[5]Facoltà di Informatica, Istituto di Scienze Computazionali, Università della Svizzera Italiana, 6900 Lugano, Switzerland

*Corresponding author: jinglu@pku.edu.cn; jbyang@pku.edu.cn

## Abstract

Absence of the Dirac cone due to a strong band hybridization is revealed to be a common feature for epitaxial silicene on metal substrates according to our first-principles calculations for silicene on Ir, Cu, Mg, Au, Pt, Al, and Ag substrates. The destroyed Dirac cone of silicene, however, can be effectively restored with linear or parabolic dispersion by intercalating alkali metal atoms between silicene and the metal substrates, offering an opportunity to study the intriguing properties of silicene without further transfer of silicene from the metal substrates.

# INTRODUCTION

As the silicon analog of graphene, free-standing silicene was predicted to have a Dirac cone in the band structure, and its electrons near the Fermi level follow the massless Dirac equation just like graphene,[1-3] which will lead to many unique properties such as ultra-high carrier mobility, anomalous quantum Hall effect, and topological insulating state.[1, 4, 5] Very recently, silicene has been fabricated by epitaxial growth on Ag,[6, 7] Ir,[8] and $ZrB_2$[9] substrates. In order to realize the predicted fascinating properties, it is crucial to maintain the Dirac cone of silicene on the substrate. So far, Ag is the most common substrate to grow silicene, and several phases have been observed for silicene on Ag substrate.[10-12] Unfortunately, no Landau level sequences is found in the scanning tunneling spectra (STS) of the (3×3)silicene/(4×4)Ag(111) and (√7×√7)silicene/(√13×√13)Ag(111) phases under a strong magnetic field, suggesting that the two Ag-supported silicene phases are neither Dirac fermion nor 2D electronic systems.[13] Subsequent theoretical calculations reveal that the Dirac cone in (3×3)silicene/(4×4)Ag(111), (√7×√7)silicene/(√13×√13)Ag(111), (√7×√7)silicene/(2√3×2√3)Ag(111), and (2×2)silicene/(√7×√7)Ag(111) phases is destructed or substantially modified as a result of the strong band hybridization between silicene and the Ag surface.[13-20] The observed linear dispersions in (3×3)silicene/(4×4)Ag(111) and (√3×√3) silicene phases are ascribed to the *s-p* bands of bulk Ag[17] or the silicene-Ag hybridization[14] instead of the intrinsic bands of silicene according to several later calculations. Although the growth of the (√3×√3) silicene phase on Ag(111) substrate is claimed by Chen *et. al* and suggested to show the Dirac cone from the STS measurement,[21] no theoretical calculation could confirm the existence of the Dirac cone in this phase (In the band structure presented in Chen's paper, the strong silicene-Ag substrate interaction is ignored.). On the contrary, the later experiment[12] and theoretical calculation[22] suggest that the (√3×√3) silicene phase on Ag(111) substrate is bilayer silicene instead of monolayer silicene, and there is also a strong silicene-Ag orbital hybridization.

It has been well established that the Dirac cone of graphene is also destroyed when chemisorbed on Ni, Co, Ti, and Pd substrates due to the significant hybridization between graphene $p_z$ and the metal *d* states, but it is preserved when physisorbed on Al, Cu, Ag, Au, and Pt substrates as a result of weak interaction.[23-26] In light of the extreme importance of the Dirac cone to silicene, the first fundamental issue arises naturally: whether the Dirac cone of silicene can be preserved on other metal substrates with a higher work function and chemical stability (e.g. Ir, Pt, and Au)?

It has been found that the intercalation of some metal atoms, such as alkali metal (Na, K, Cs),[27, 28] *sp*-metal (Al),[29, 30] and noble metal (Au and Cu)[31-34] into the graphene/Ni(111) interface can effectively weaken the strong interaction between graphene and the underlying Ni substrate and restore the destroyed Dirac cone. The recovered Dirac cone of graphene is located exactly at the Fermi level ($E_f$) with an Au intercalated layer, but below $E_f$ when intercalated with other metal atoms mentioned above. Back to silicene science, the second fundamental issue is whether the destroyed Dirac cone of silicene on metal substrates can be recovered in the intercalation way.

In this Article, by using first-principles calculations, we studied the structural and electronic properties of silicene on different metal substrates, including Ir, Au, Pt, Al, Cu, Mg, and Ag. It is found that on Ir, Au and Pt substrates, silicene is *p*-type doped, whereas on Al, Cu, and Mg substrates, silicene is *n*-type doped. The band hybridization between silicene and all the examined metals turns out to be rather strong without exception, resulting in the severe destruction of the Dirac cone of silicene. However, by virtue of the alkali metal atom intercalation between silicene and substrates, both the massless (on Au, Pt and Al substrates) and massive (on Ag, Ir, Cu, and Mg substrates) Dirac fermions can be restored in silicene.

## MODEL AND METHODS

We put one layer of ($\sqrt{3}\times\sqrt{3}$) silicene on top of metal surfaces including Ir(111), Au(111), Pt(111), Cu(111), Al(111), and Mg(0001), and (2×2) silicene on top of Ag(111) surface. The atomic arrangement of the metal surface is a ($\sqrt{7}\times\sqrt{7}$) superlattice expect for Mg with a (2×2) superlattice. In the intercalation model, a layer of alkali metal atoms is inserted between silicene and the metal surface (the ratio of the number of Si atoms to that of alkali metal atoms is 2). The ($\sqrt{3}\times\sqrt{3}$)silicene/($\sqrt{7}\times\sqrt{7}$)Ir(111)[8] and (2×2)silicene/($\sqrt{7}\times\sqrt{7}$)Ag(111)[10, 12] configurations have already been observed by scanning tunneling microscopy. There are five layers of metal atoms in each slab. Five layers of metal atom have also been used in previous silicene/Ag interface models, [13, 16, 17, 19] and the resulting electronic properties, especially the strong Si-Ag band hybridization, show no significant difference with other model using more Ag atom layers. [14, 15] We adapt the lattice constant of the silicene layer to the in-plane lattice constant of bulk metals (Table 1). The mismatches of silicene with the lattice parameters of the metal surfaces $\Delta a$ are 0.99 ~ 14.95%. A vacuum space of at least 15 Å is applied in the $z$ direction. In the energy barrier calculations of an alkali metal atom penetrating through silicene, a (5×5) silicene supercell is adopted.

Geometry optimization and electronic properties are calculated by using an ultrasoft pseudopotential and plane-wave basis set with a cut-off energy of 400 eV, as implemented in CASTEP.[35] The geometry optimization is performed until the remaining forces become less than $10^{-2}$ eV/Å. The generalized gradient approximation (GGA)[36] of the Perdew-Burke-Ernzerhof (PBE) exchange-correlation functional is adopted. Van der Waals interactions are taken into account (PBE–D) using an approach by Tkatchenko and Scheffler.[37] The Monkhorst-Pack $k$-point mesh[38] is sampled with a separation of about 0.02 Å$^{-1}$ in the Brillouin zone. Since the slab is not symmetric, a dipole correction is used to eliminate the spurious interaction between the dipole moments of periodic images in the $z$ direction. Metal atoms in the bottom three layers are kept fixed at bulk lattice positions during geometry optimization. The charge transfer is computed by using Mulliken population analysis.[39] The linear synchronous transit (LST) method,[40] followed by an energy minimization, is applied to determine penetration pathways. The component and wave function of the energy band are analyzed with resort to additional calculations based on the plane-wave basis set with a cut-off energy of 350 eV and the projector-augmented wave (PAW) pseudopotential,[41] implemented in the Vienna *ab initio* simulation package.[42, 43] The whole electronic band structure (for silicene on Mg with K intercalation) calculated by CASTEP with the ultrasoft pseudopotential is almost the same as that calculated by VASP with PAW pseudopotential (Fig. S1).

## RESULTS AND DISCUSSION

The key interfacial structure and property parameters of silicene on metal substrates are summarized in Table 1. The silicene-metal interfacial structures can be classified into two categories. Interface I includes Ir, Mg, Cu, and Ag substrates, where the Si atoms are distributed on two (Ir, Mg, Cu) or three (Ag) different heights, with much larger buckling heights ($\Delta$ = 0.83 (Ir), 1.52 (Mg), 1.25 (Cu), and 1.16 Å (Ag)) than that (0.46 Å) in free-standing silicene. In each ($\sqrt{3}\times\sqrt{3}$) or (2×2) silicene supercell, only one Si atom is located right above the center of the metal atom with a larger height (Fig. 1a), and the other Si atoms are located on either the metal-metal bonds or the hollow centers among three metal atoms. The calculated buckling and the distance ($d_0$ = 1.95 Å) between silicene and Ir substrate are in good agreement with other calculations.[8] Interface II includes Pt, Al, and Au substrates, where the Si atoms are distributed on two close heights, with $\Delta$ = 0.33, 0.29 and 0.21 Å, respectively (Fig. 1b). The small values of buckling in Interface II may be related to the large lattice mismatches $\Delta a$ (greater than 9%) between silicene and these metal substrates. The silicene-metal distances $d_0$ on these metals are less than 2.3 Å, similar to the graphene-metal distances of graphene chemisorbed on Ni, Co, and Pd(111) substrates.[23, 24]

The binding energy $E_b$ of the epitaxial silicene on metal substrates is defined as below:

$$E_b = (E_{Si} + E_M - E_{Si/M})/N \qquad (1)$$

where $E_{Si}$, $E_M$, and $E_{Si/M}$ are the energy for free-standing silicene layer, clean metal substrates, and composite systems, respectively, and $N$ is the number of Si atoms per supercell. The calculated $E_b$ of silicene on Ir substrate is 1.69 eV/Si atom, in good agreement with previous calculations.[8] Compared to Ir substrate, silicene bonds relatively weakly to the Al, Mg, Ag, Au, and Cu substrates with $E_b$ = 0.35, 0.39, 0.41, 0.63, and 0.86 eV/Si atom, respectively, while more strongly to Pt substrates with $E_b$ = 1.98 eV/Si atom. The interactions between silicene and metal substrates are much stronger than those between graphene and metal substrates ($E_b \sim$ 0.1 eV/C atom for physisorption on Ag, Cu, Au, Ir, Pt, and Al substrates and $E_b \sim$ 0.2 eV/C atom for chemisorption on Ni and Co substrates calculated at the same PBE−D level).[26]

The formation energy ($G$) of epitaxial silicene on metal with bulk Si as the reservoir of Si has been computed. As shown in Table I, the $G$ for epitaxial silicene on Pt, Ir, Cu, and Au are positive, while that on Ag, Mg, and Al substrates are negative. The relative stability of the epitaxial silicene on different metal substrates ascends in the following order: Al < Mg < Ag <

Au < Cu < Ir < Pt. According to the calculated formation energy, the epitaxial silicene on Pt, Ir, Cu, and Au substrates are more stable than on the commonly used Ag substrate.

The band structures of epitaxial silicene on various metal substrates are shown in Fig. 2, where the red dots stand for the bands contributed by the Si atoms. We can hardly see the original "cone" shape band structure of free-standing silicene in all the band structures, suggesting a strong band hybridization between silicene and the examined metal substrates. The near-cone shape in Fig. 2c (Au substrate) originates from the $s$ as well as the $p_x$ and $p_y$ orbitals of Si atoms, instead of the $p_z$ orbital that contributes to the Dirac cone of silicene. Compared to other substrates, more states from silicon atoms are found in the band structure of silicene on Mg substrate as shown in Fig. 2d. However, the "cone" shape bands are still completely destroyed. The inset in Fig. 2f (Cu substrate) is the electron charge density at the Γ point denoted by the black square. Even though the state stems mainly from the Si atoms, there is still contribution from the Cu substrate.

The difference between the work function ($W$) of the graphene-covered metal substrate and free-standing graphene has been used to describe the doping level of graphene chemisorbed on the metal substrate approximately, where the Dirac cone is severely destroyed.[23, 24, 26] Similarly, we take this scheme to describe the doping level of the epitaxial silicene chemisorbed on metal substrates. The work functions $W$ of epitaxial silicene on Ir, Au, and Pt substrates are 0.57, 0.08, and 0.07 eV greater than that of free-standing silicene (4.48 eV), and silicene loses 0.13, 0.06, and 0.04 electrons per Si, respectively, suggesting a heavy $p$-type doping of silicene by Ir substrate and a low $p$-type doping of silicene by Au and Pt substrates. Together with the band structure, we can conclude that the interaction between silicene and metal substrates is a mixture of covalent and ionic bonds. In contrast, the work functions $W$ of epitaxial silicene on Mg, Al, Ag, and Cu substrates are 0.50, 0.36, 0.11, 0.03 eV less than that of free-standing silicene, and silicene obtains 0.21, 0.06, 0.07, and 0.04 electrons per Si atom, respectively, suggesting a heavy $n$-type doping of silicene by Mg and Al substrates and a relatively low $n$-type doping of silicene by Ag and Cu substrates. Mono-, bi-, and tri-layer graphene, which share similar work function values (4.48, 4.58, and 4.52 eV, respectively) with silicene, are also $p$-doped on Ir, Pt, and Au substrates while $n$-doped on Al, Cu, and Ag substrates.[23, 24, 26]

The Si coverage of the considered (2×2) silicene on (√7×√7) Ag configuration is 1.143.[18] Other combinations of translational symmetry ((3×3)silicene/(4×4)Ag(111), (√7×√7)silicene/(√13×√13)Ag(111), and (√7×√7)silicene/(2√3×2√3)Ag(111) phases) have reached the same conclusion that the Dirac cone is absent for silicene due to the covalent interaction between silicene and Ag substrates based on a variety of calculations.[13-15, 17, 19, 20] Though we only choose one configuration for each interface and even the lattice mismatch between silicene and Pt (9.6%), Al (13.2%), and Au (14.0%) is large, the conclusion of the absence of the Dirac cone for silicene on these metal substrates due to the strong band hybridization feature between Si and these metal substrates should be unaffected by using other models with a smaller lattice mismatch.

Because Ir, Pt, and Au are among the metals with the highest work function and highest chemical stability, it is unlikely to avoid the band hybridization with silicene using other metal substrates. In order to recover the Dirac cone destructed by the strong band hybridization between silicene and metal surface, we introduce the intercalation of alkali metal atom scheme. The intercalated alkali metal atoms prefer being located beneath the hollow center of the Si honeycomb, as shown in Figs. 1c and 1d. The key interfacial structure and electronic property parameters of the alkali-intercalated silicene on metal substrates are summarized in Table 2. The structures of silicene in Interfaces I and II become quite different after the intercalation of alkali metal atoms. The Si atom layer in K-intercalated Interface I (Ir, Mg, Cu, and Ag) shows the same every second up-every second down structure as that of free-standing silicene, and the buckling values range from 0.37 ~ 0.61 Å. However, the Si atoms in K-intercalated Interface II (Pt, Al, and Au) are almost in the same plane. Considering the large lattice mismatch (> 9%) of Interface II, we have optimized the free-standing silicene under tensile strength from 8% to 14% and found that the buckling of silicene is still kept. Therefore, the disappearance of buckling of silicene in Interface II is not solely due to the strong tensile strength but also related to the alkali metal atom intercalation.

The calculated separation distance of silicene from alkali metal atoms $d_1$ and that of alkali metal atoms from metal substrates $d_2$ range from 2.51 to 3.17 Å. The energy required to remove the silicene sheet from the alkali metal atom-adsorbed metal surface $E_b'$ is defined as below:

$$E_b' = (E_{Si} + E_{AM/M} - E_{Si/AM/M})/N \qquad (2)$$

where $E_{Si}$, $E_{AM/M}$ and $E_{Si/AM/M}$ are the energy for free-standing silicene, alkali-adsorbed metal substrates, and composite systems, respectively, and $N$ is the number of Si atoms per supercell. The values of $E_b'$ range from 0.39 to 0.86 eV/Si atom. Compared with the case without intercalation of K atoms, the binding energy increases by 0.07 ~ 0.46 eV/Si atom for silicene on Ag, Mg, and Al substrates, while decreases by 0.01 ~ 0.99 eV/Si atom for silicene on Ir, Pt, Au, and Cu substrates.

The intercalation energy of alkali metal atoms penetrating into the space between silicene and the metal substrate is defined as below:

$$E_i = (m \times E_{AM} + E_{Si/M} - E_{Si/AM/M})/N \qquad (3)$$

where $E_{AM}$, $E_{Si/M}$ and $E_{Si/AM/M}$ are the energy for isolated alkali metal atoms, epitaxial silicene on metal substrates, and alkali atoms intercalated silicene on substrates, respectively, and $m$ and $N$ are the number of alkali metal atoms and Si atoms per supercell, respectively. The positive $E_i$ values (2.92 ~ 4.55 eV) in all the examined cases indicate that the process of alkali metal atoms penetrating through the epitaxial silicene layer is strongly exothermic.

Due to the limitation of computational resources, we have calculated the energy barrier of an alkali metal atom to penetrate free-standing silicene to estimate the barrier of the alkali metal to penetrate silicene on metal substrates. Duan *et. al.* also replaced epitaxial graphene on substrate with free-standing graphene to study the barrier height of the alkali atom intercalation,[44] and the calculated results agreed well with the experiment.[45] The calculations of the minimum-energy path gives an energy barrier of as high as 8.59 eV for a K atom penetrating through the free-standing silicene, as shown in Fig. 3a. A large energy barrier has also been calculated for a Li atom penetration through perfect graphene on SiC substrate (3.98 eV), but experimentally such a penetration can occur.[45] A defect-mediated intercalation mechanism has been proposed to account for alkali metal atoms penetration through graphene on SiC substrate and Pb atoms penetration through graphene on Ru substrate.[46] Actually, introducing point defects (heptagon and octagon) in graphene leads to a significant decrease of the energy barrier of a Li atom penetration by 2 ~ 3 times.[44, 47]

We consider four types of defect in silicene: Stone–Wales defect (SW), single (SV), double vacancies (DV), and silicene nanomesh. In the nanomesh configuration, one hexagon ring consisting six Si atoms is removed per (5×5) silicene supercell. The dangling Si atoms in the edge of the holes in DV and nanomesh configurations are hydrogenated. We find that the energy barriers for K penetrating through silicene with SW and SV defects decrease to 4.94 and 3.84 eV, respectively. Replacing K with smaller Na atoms, the energy barriers for penetrating through free-standing and SW-defected silicene decrease by 50% and 87% to 4.31 and 0.62 eV, respectively, as shown in Fig. 3b. Therefore, Na can penetrate SW-defected silicene.[48, 49] In the cases of the K penetrating the larger-sized DV and K/Na penetrating the silicene nanomesh, the energy barriers disappear, suggesting a much easier penetration. Therefore, large point defect or small size of alkali metal atoms is favorable for penetration of alkali metal atoms through the silicene layer. Moreover, stretch is found to be able to decrease the penetration barrier too. After a 7.1% stretching to free-standing silicene is applied, the penetration barrier for a Na atom decreases from 4.31 to 1.79 eV.

To study the electronic properties of alkali-metal-intercalated Interfaces I (Ag, Mg, Cu, and Ir) and II (Pt, Al, and Au), we first look at the band structures of silicene with the intercalated K atoms and Ag/Pt substrate removed (we refer to them as unsupported silicene I and II, respectively). As shown in the green color of Fig. 4a, the bands of unsupported silicene I are quite similar to that of free-standing silicene, with the Dirac cone located at $E_f$. In the band structure of unsupported silicene II (Fig. 5a), the Dirac cone lies ~ 0.3 eV above $E_f$. Meanwhile, there is a 3s band of Si atom across the Dirac cone (green color). In fact, the position of the σ states at Γ point relative to $E_f$ in silicene is very sensitive to the tensile biaxial strain and buckling.[50, 51] This phenomenon has also been observed in the single Ge atom layer: the σ states also goes below $E_f$ in the Γ point for the planer form but above $E_f$ for the low buckled form.[52] It appears that under circumstances of applying tensile biaxial strain on flat systems, the σ states at the Γ point tend to approach $E_f$.

After intercalation of K atoms, the Dirac cone of silicene is recovered in both Interfaces I and II, and the general shape of the Si-originated bands is quite similar to that of corresponding unsupported silicene. The recovered Dirac cone is located at 0.40 ~ 0.78 eV below $E_f$, suggesting an n-type doping of silicene. However, comparing the band structures of K-intercalated Interfaces I and II in Figs. 4 and 5, two important differences are noteworthy. First, in K-intercalated Interface I, a band gap of 0.15 ~ 0.40 eV is opened between the π and $\pi^*$ bands, as a result of the breaking of inversion symmetry[53, 54] between the two silicene sublattices in different planes and intervalley interaction.[54-58] The breaking of the sublattice symmetry is due to the build-in electric field vertical to the silicene plane induced by the charge transfer. The intervalley interaction works because the two valleys K and K' were manipulated to the Γ point at the (√3×√3) Si supercell, except for the (2×2) Si/(√7×√7) Ag configuration in Fig. 4a . The size of the band gap is comparable with that of K adsorbed silicene with the coverage of 5.6 ~ 50%.[54] The energy dispersion around the Dirac cone is parabolic, suggesting a massive fermion. In K-intercalated Interface II (Fig. 5), no band gap is opened between the $\pi$ and $\pi^*$ bands, even though the degeneracy of the four states at the Dirac point in unsupported silicene II is slightly broken. The degeneracy of the conduction and valence band at the Γ point because the two sublattices of silicene are in the same plane and the inversion symmetry between them is recovered. Moreover, there is no intervalley interaction at the Γ point because the alkali atoms are uniformly distributed below every honeycomb of the flat silicene layer (degenerate to a (1×1) silicene).

The energy dispersion near the Dirac point is near-linear, suggesting a near-massless fermion. The estimated values of the Fermi velocity of K-intercalated Interface II are $4.84×10^5$, $4.45×10^5$, and $4.20×10^5$ m/s for Pt, Al, and Au substrates, respectively, only slightly less than the value of $6.15×10^5$ m/s in free-standing silicene calculated at the same level. Second, in K-intercalated Interface II, there is overlap between the *s* band of Si atoms and the Dirac point in energy. This will increase the difficulty to observe the massless Dirac fermion by measuring the Landau-level separation.

We have calculated the band structures of the silicene layer with K atoms in the bottom but the Cu or Ir substrate removed and show them in Figs. 4e and 4f, respectively. The resulting band structures are in good agreement with that of the previous report of K-adsorbed silicene with the same coverage[54]: the Dirac cone with a band gap (0.3 ~ 0.5 eV) is observed, and a K-derived band appears, which is below $E_f$ near the Γ point and is lower in the present work. Comparing the band structures with and without substrates, the bands derived from Si atoms including the position of the Dirac cone with respect to $E_f$ are almost the same. We conclude that the recovering of the Dirac cone is due "only" to the presence of the alkali atoms, irrespective of the substrate.

The electron distributions at the Dirac point for free-standing silicene, unsupported silicene, and silicene in K-intercalated Interfaces I and II are compared in Fig. 6. The electron densities of the $\pi$ states of free-standing silicene (Fig. 6a), unsupported silicene I (Fig. 6b), and silicene in K-intercalated Interface I (Fig. 6c) are in remarkable resemblance. The electron densities of the $\pi^*$ state in the latter two cases are even nearly the same. The difference of the $\pi^*$ state between free-standing and unsupported silicene is attributed to the strain in unsupported silicene. The electron densities for States 1-2 are similar among free-standing plane silicene (Fig. 6d), unsupported silicene II (Fig. 6e), and silicene in K-intercalated Interface II (Fig. 6f). High similarity is also observed for the electron densities of State 3 between unsupported silicene II (Fig. 6e) and silicene in K-intercalated Interface II (Fig. 6f) and for the electron densities of State 4 between free-standing plane silicene (Fig. 6d) and silicene in K-intercalated Interface II (Fig. 6f). The difference of State 3 between free-standing plane silicene and unsupported silicene also can be attributed to the strain difference. The general existence of similarity of the electron densities at the Dirac point between free-standing silicene, unsupported silicene, and silicene in K-intercalated silicene/metal interfaces further confirm that the Dirac cone of silicene on metal substrates is recovered or partially recovered upon alkali metal atom intercalation.

The total electron distributions in real space without and with K intercalation are provided in Fig. 7. Before the intercalation of K atoms (Figs. 7a and 7c), the electrons are distributed not only around the Si and Ir (Au) atoms but also between the silicene layer and the metal surfaces, indicating the formation of the covalent bond between silicene and the metal substrate in addition to the formation of the ionic bond caused by charge transfer. However, after the intercalation of K atoms, electrons prefer being localized around the Si and K atoms, suggesting a dominated ionic bond. The change of the interaction between silicene and metal surface/alkali metal atoms from the mixture of the covalent and ionic bond to pure ionic covalent is the exact reason why the Dirac cone in metal-supported silicene can be recovered by intercalation of alkali metal atoms. The covalent Si-metal bond causes strong band hybridization between silicene and metal and thus deformation or even disruption of the Dirac cone, whereas the ionic Si-metal bond merely causes a rigid shift of the Dirac cone of silicene.

As a side remark, we have checked the dependence of the geometrical and electronic structure on the type of the intercalated alkali metal atoms in epitaxial silicene on Ir substrate. The epitaxial silicene on Ir with Li intercalation is not stable, and the small Li atoms penetrate silicene and move above the silicene layer after optimization. Therefore, the structural parameters of this configuration are not provided, and strong band hybridization of silicene and Ir substrate still exists in the band structure as shown in Fig. S2a. With Na-intercalation, the buckling of silicene on Ir is as the same as that of free-standing silicene and decreases by 0.09 and 0.14 Å with K- and Rb-intercalation, respectively. The equilibrium silicene-alkali distance $d_1$ and alkali-Ir surface distance $d_2$ are positively related to the atomic radius of the intercalated alkali metal atoms, raising form $d_1 = 2.55$ Å and $d_2 = 2.51$ Å to $d_1 = 2.96$ Å and $d_2 = 3.04$ Å as the atomic radius of alkali metal atoms increases (Table 2). While the structural parameters of silicene on Ir with alkali intercalation are strongly related to the type of alkali metal atoms, their electronic structures are quite similar. The recovered Dirac cone by Na/K/Rb intercalation is located at ~ 0.6 eV below $E_f$ with a direct $\pi$-$\pi^*$ band gap at the Γ point, as shown in Figs. S2b-2d.

The intercalation of atoms like Au, K and Cu is able to recover the destroyed Dirac cone of graphene on Ni(111)[31, 33, 59, 60] due to the ionic bond character between Au/K/Cu atoms and graphene[61]. However, silicene is more reactive than graphene, the interaction between Au/Cu atoms has a partial covalent bond character at a high coverage, and the Dirac cone of silicene is seriously disturbed.[58] Therefore it is unlikely to recover the Dirac cone of silicene by intercalation of high-concentration Au and Cu atoms between silicene/metal interfaces.

## CONCLUSION

In conclusion, unlike the case of graphene, the absence of the Dirac cone due to an enhanced band hybridization appears to be a common character for silicene on metal substrates from our first principles calculations of silicene on a series of metal substrates. However, by intercalation of alkali metal atoms between silicene and metal substrates, the Dirac cone of silicene can be recovered with linear or parabolic dispersion. Although Au[62, 63], Ag, Cu, and Ir substrates have a strong band hybridization with silicene on them, atom adsorption of the four metals on silicene with a low coverage can keep the Dirac cone of silicene.[58] Therefore, intercalating low-concentration other metal atoms with a higher work function, such as Au, may result in a recovered Dirac cone near or above the Fermi level and is worthy to be further explored. We note that the Dirac cone of silicene on $ZrB_2$ substrate is also destroyed [64] by the interaction between silicene and $ZrB_2$ substrate. We expect that the destroyed Dirac cone of silicene on $ZrB_2$ substrate can also be recovered by intercalation of alkali metal atoms between silicene and $ZrB_2$ substrate.

**Acknowledgement**

This work was supported by the National Natural Science Foundation of China (Nos. 11274016, 51072007, 91021017, 11047018 and 60890193), the National Basic Research Program of China (Nos. 2013CB932604 and 2012CB619304), Fundamental Research Funds for the Central Universities, National Foundation for Fostering Talents of Basic Science (No. J1030310/No. J1103205), Program for New Century Excellent Talents in University of MOE of China. R. Quhe also acknowledges the financial support from the China Scholarship Council. J. Lu and R. Quhe thank Prof. K. H. Wu and Prof. M. Parrinello for helpful discussions. Thanks are also due to Dr. J. J. McCarty V and Dr. P. Tiwary for their help in revising the English language.

**Additional Information**

**Contributions**

The idea was conceived by J. L. The calculation was performed by R. Q., Y. Y., Y. W., Z. N., and J. Z. The data analyses were performed by J. L., J. Y., D. Y., J. S., and R. Q. This manuscript was written by R. Q., Y. Y., J. Y., and J. L. All authors contributed to the preparation of this manuscript.

**Competing financial interests**

The authors declare no competing financial interests.

**Table 1.** Structural parameters of the epitaxial silicene on different metal substrates. The structural phase is shown in the first row. $a$ and $a'$ are the lattice constants of metal substrates and corresponding free-standing silicene in a supercell, and $\Delta a$ is the lattice mismatch between metal and free-standing silicene. $\Delta$ is the calculated buckling of silicene. $d_0$ is the equilibrium separation of silicene from various metal surfaces. The binding energy $E_b$ is the energy (per Si atom) required to remove the silicene sheet from the metal surface. $G$ is the formation energy (per Si atom) of epitaxial silicene on metal with bulk Si as the reservoir of Si. $W_m$ and $W$ are the work functions for the clean metal surface and for the metal-supported silicene, respectively. $Q$ is the Mulliken charge per Si atom transferred from silicene to the metal surfaces. The work function of free-standing silicene is 4.48 eV.

| | ($\sqrt{3}\times\sqrt{3}$)Si/ ($\sqrt{7}\times\sqrt{7}$)Ir | | ($\sqrt{3}\times\sqrt{3}$)Si/ (2×2)Mg | ($\sqrt{3}\times\sqrt{3}$)Si/ ($\sqrt{7}\times\sqrt{7}$) Cu | (2×2)Si/($\sqrt{7}\times\sqrt{7}$)Ag[b] | ($\sqrt{3}\times\sqrt{3}$)Si/ ($\sqrt{7}\times\sqrt{7}$)Pt | ($\sqrt{3}\times\sqrt{3}$)Si/ ($\sqrt{7}\times\sqrt{7}$)Al | ($\sqrt{3}\times\sqrt{3}$)Si/ ($\sqrt{7}\times\sqrt{7}$)Au |
|---|---|---|---|---|---|---|---|---|
| $a$ (Å) | 7.18 | 7.2[a] | 6.42 | 6.76 | 7.64 | 7.34 | 7.58 | 7.63 |
| $a'$ (Å) | 6.70 | | | 6.70 | 6.70 | 7.73 | 6.70 | 6.70 | 6.70 |
| $\Delta a$ (%) | 7.2 | | -4.1 | 1.0 | -1.2 | 9.6 | 13.2 | 14.0 |
| $\Delta$ (Å) | 0.83 | 0.83[a] | 1.52 | 1.25 | 1.16 | 0.33 | 0.29 | 0.21 |
| $d_0$ (Å) | 1.95 | 2[a] | 2.28 | 1.57 | 1.39 | 1.97 | 2.12 | 1.81 |
| $E_b$ (eV) | 1.69 | 1.6[a] | 0.39 | 0.86 | 0.41 | 1.74 | 0.35 | 0.63 |
| $G$ (eV) | 1.10 | | -0.20 | 0.27 | -0.19 | 1.15 | -0.24 | 0.04 |
| $W_m$ (eV) | 5.47 | | 3.55 | 4.69 | 4.46 | 5.82 | 4.06 | 5.09 |
| $W$ (eV) | 5.05 | | 3.98 | 4.45 | 4.36 | 4.55 | 4.12 | 4.56 |
| $Q$ ($|e|$) | 0.13 | | -0.21 | -0.04 | -0.07 | 0.06 | -0.06 | 0.04 |

[a] Reference [8].

[b] Reference [20].

**Table 2.** Calculated buckling of silicene ($\Delta$), equilibrium separation of silicene from alkali metal atoms ($d_1$) and alkali metal atoms from various metal surfaces ($d_2$). Na, K and Rb atom intercalations are considered in the case of Ir(111), and only intercalation of K atoms is considered for the other metal surfaces. The binding energy $E_b'$ is the energy (per Si atom) required to remove the silicene sheet from the alkali metal atom-adsorbed metal surface. The intercalation energy $E_i$ is the energy (per Si atom) required to insert alkali metal atoms between epitaxial silicene and metal substrates. The Fermi level shift $\Delta E_f$ is defined as $\Delta E_f = E_D - E_f$.

|  | Ir | | | Mg | Cu | Ag | Pt | Al | Au |
|---|---|---|---|---|---|---|---|---|---|
|  | (Na) | (K) | (Rb) | | | | | | |
| $\Delta$ (Å) | 0.46 | 0.37 | 0.32 | 0.54 | 0.61 | 0.50 | 0 | 0 | 0 |
| $d_1$ (Å) | 2.55 | 2.93 | 2.96 | 2.62 | 2.81 | 2.68 | 3.05 | 2.88 | 2.80 |
| $d_2$ (Å) | 2.51 | 2.89 | 3.04 | 3.17 | 2.68 | 2.93 | 2.77 | 3.05 | 2.76 |
| $E_b'$ (eV) | 0.56 | 0.70 | 0.72 | 0.82 | 0.85 | 0.86 | 0.81 | 0.42 | 0.39 |
| $E_i$ (eV) | 2.92 | 3.31 | 3.37 | 3.05 | 4.55 | 4.03 | 3.19 | 4.14 | 3.30 |
| $\Delta E_f$ (eV) | -0.60 | -0.62 | -0.60 | -0.78 | -0.52 | -0.75 | -0.50 | -0.50 | -0.40 |

**Figure 1.** (a-b) Optimized structures of epitaxial (√3×√3) silicene on (√7×√7) Ir and (√7×√7) Pt substrates. (c-d) Optimized structures of epitaxial (√3×√3) silicene on (√7×√7) Ir and (√7×√7) Pt substrates after intercalation of K atoms. Yellow (red), blue and green balls are Si, Ir and Pt atoms, respectively. The Si atoms in red in (a) has a longer distance to the Ir surface than the other Si atoms (yellow).

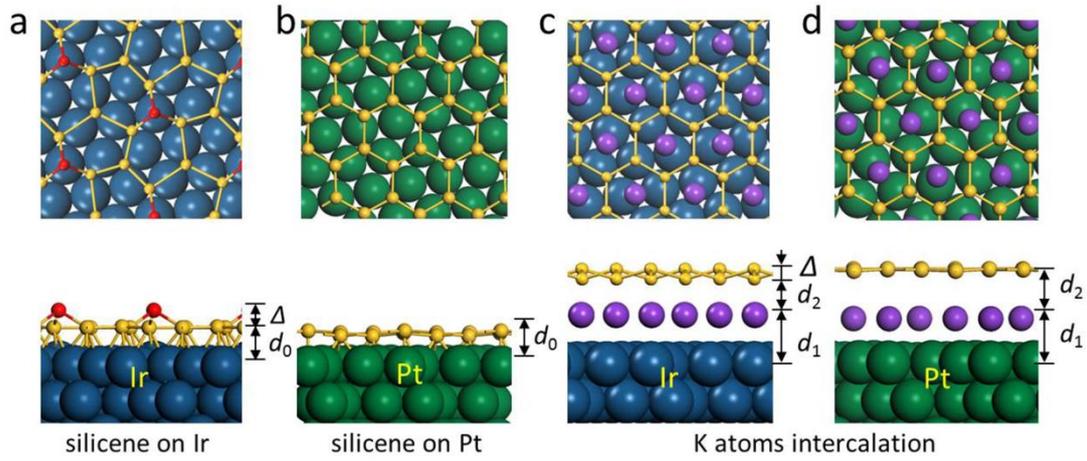

**Figure 2.** Band structures of epitaxial (√3×√3) silicene on various metal surfaces. The Dirac point is folded at the Γ point. The Fermi level is set to zero. The red color in (a-b) and (d-f) indicates the states contributed by the Si atoms, and the green, blue, and magenta colors in (c) indicate the states contributed by $s$, $p_x$ and $p_y$, and $p_z$ orbitals of the Si atoms, respectively. The thickness of these colors is proportional to the Si atom character. The inset in (f) is the electron density at the Γ point inside the black square, and the yellow and silver balls are Si and Cu atoms, respectively.

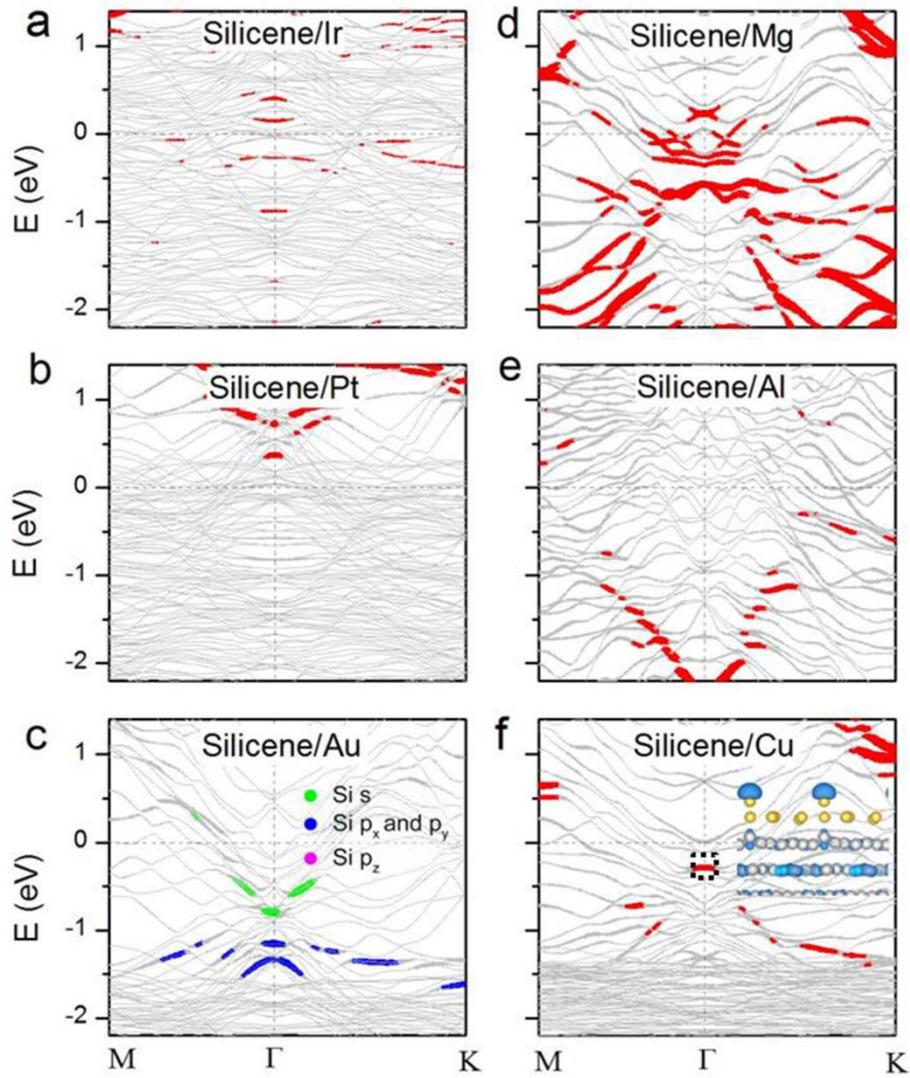

**Figure 3.** Calculated energy paths of (a) K and (b) Na atom penetrating through the free-standing, stretched silicene, silicene with point defects, and silicene nanomesh. The energy of the initial state is set to zero. The insets from left to right are the configurations of single (SV), and double vacancies (DV), Stone–Wales defect (SW), and silicene nanomesh. In the nanomesh configuration, one hexagon ring is removed per (5×5) silicene supercell. The dangling Si atoms in the edge of the holes in DV and nanomesh configurations are hydrogenated. The yellow and white balls are Si and H atoms, respectively.

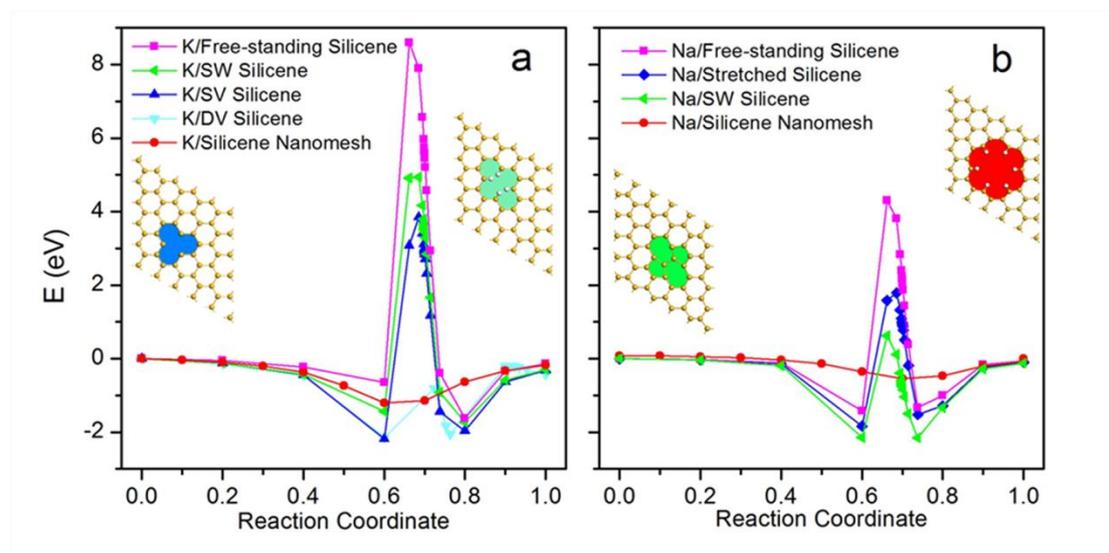

**Figure 4.** Band structures of epitaxial silicene on various metal substrates with K atoms intercalation: (a) (2×2) silicene on (√7×√7) Ag substrate, (b-d) (√3×√3) silicene on (2×2) Mg, (√7×√7) Cu, and (√7×√7) Ir substrates, (e-f) K-intercalated (√3×√3) silicene with Cu and Ir substrates removed. The Fermi level is set to zero. The red color indicates the states contributed by the Si atoms, and its thickness is proportional to the Si atom character. The green lines in (a) are the band structure of the epitaxial silicene with the intercalated K atoms and Ag substrate removed. The Dirac point in (b-d) is folded to the Γ point. Black squares indicate those states contributed by the $\pi$ ($\pi^*$) bands of silicene.

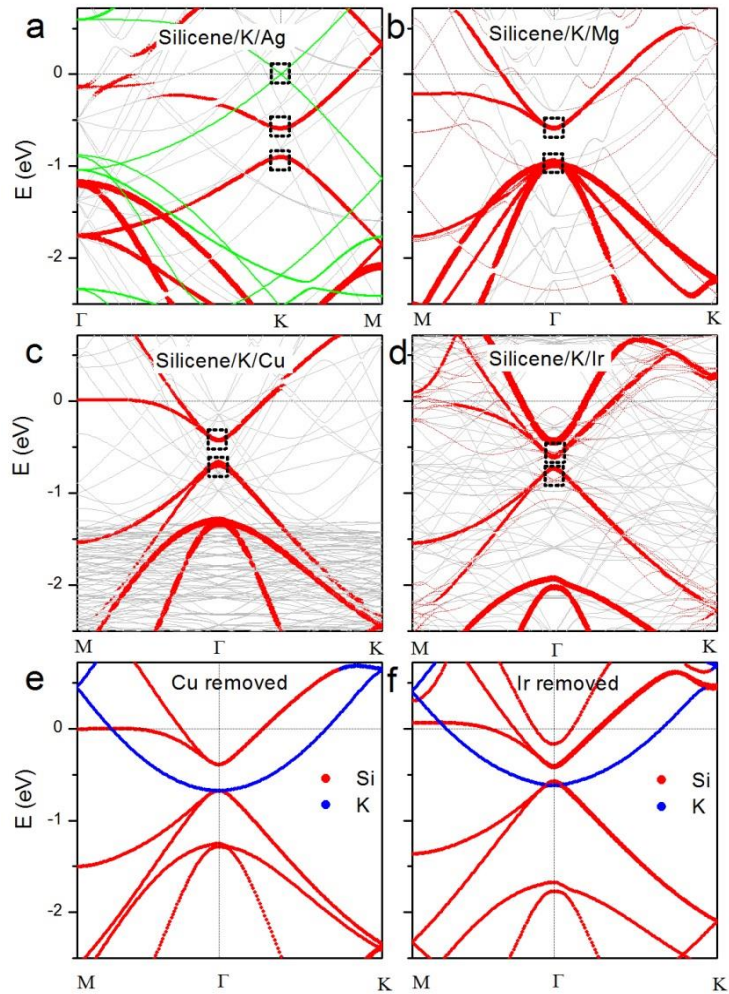

**Figure 5.** (a) Band structure of epitaxial (√3×√3) silicene with the intercalated K atoms and the Pt substrate removed. (b-d) Band structures of epitaxial silicene on Pt, Al, and Au substrates with K atoms intercalation. The Fermi level is set to zero. The green, blue, and magenta colors in (a) indicate the states contributed by $s$, $p_x$ and $p_y$, and $p_z$ orbitals of the Si atoms respectively, and the red color in (b-d) indicates the states contributed by the Si atoms. The thickness of these colors is proportional to the Si atom character. The Dirac point is folded to the Γ point and indicated by a square.

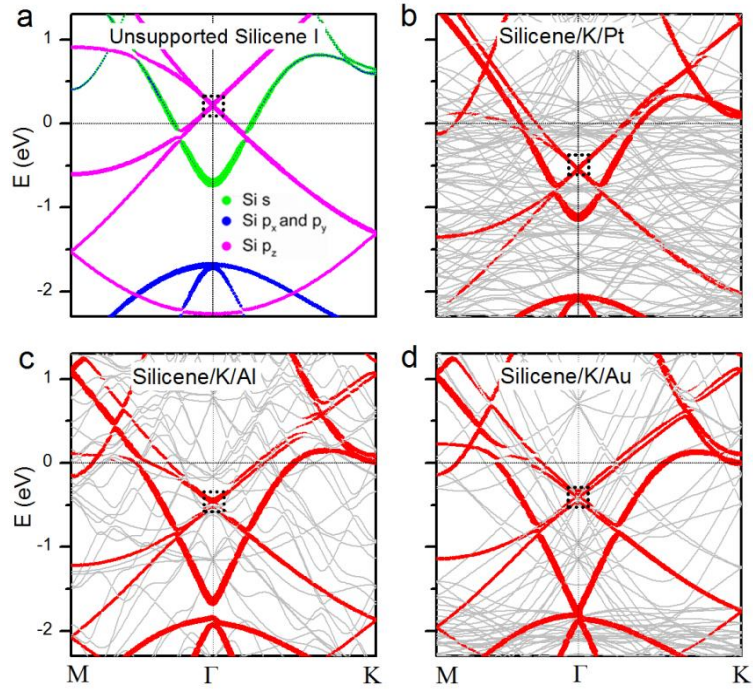

**Figure 6**. Isosurfaces of electron density at the Dirac point: (a-c) the $\pi$ and $\pi^*$ states of free-standing silicene (a), and silicene without (b) and with (c) K atoms and Ag substrate, and (d-f) the four mixed $\pi$ ($\pi^*$) states of free-standing ($\sqrt{3} \times \sqrt{3}$) plane silicene (d), and silicene without (e) and with (f) K atoms and Pt substrate. Yellow, purple, silver, and green balls are Si, K, Ag, and Pt atoms, respectively. The isovalue is 0.02 $e$/Å$^3$.

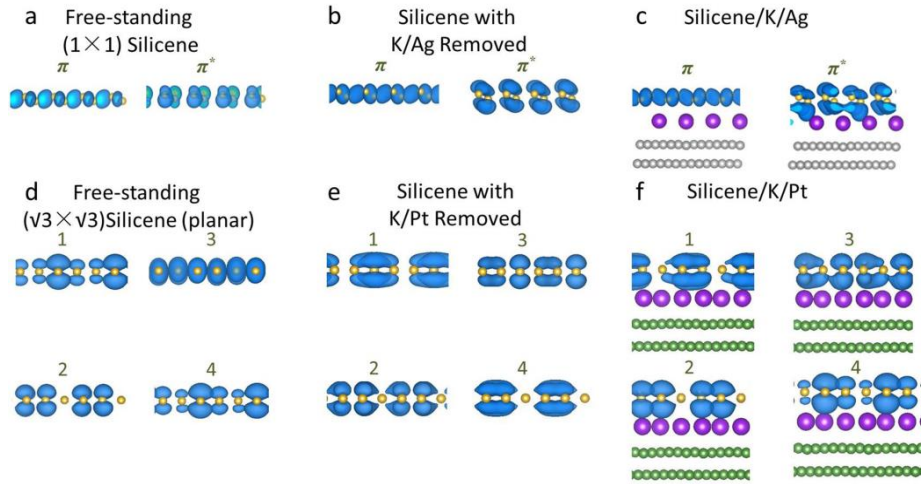

**Figure 7.** Contour plots of total electron distribution of (a-b) Interfaces I: silicene/Ir and (c-d) II: silicene/Au before and after intercalation of K atoms. Yellow, blue, and green balls are Si, Ir, and Au atoms, respectively.

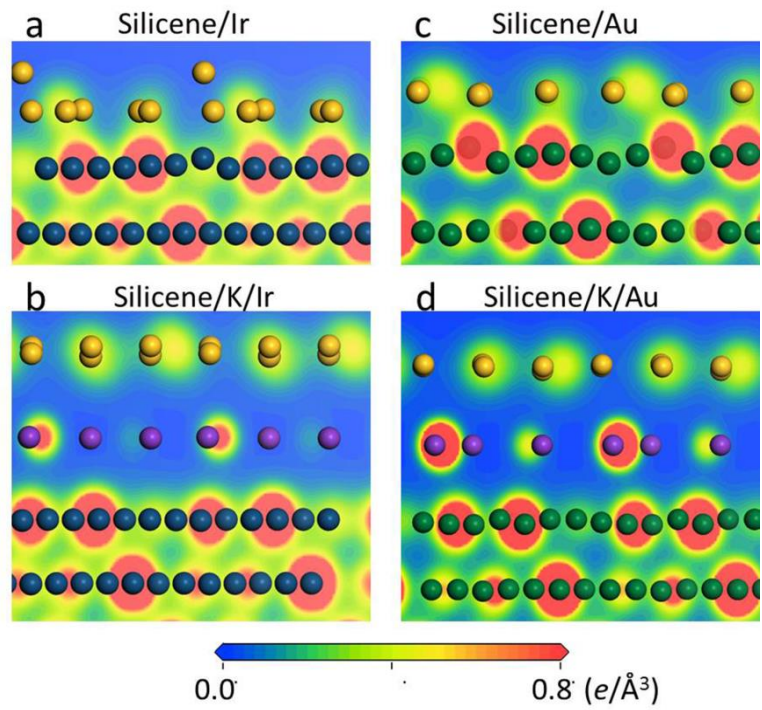